\documentstyle[epsf,epsfig,preprint,aps,amssymb]{revtex}
\draft \preprint{SNUTP 01/034}
\begin{document}
\title{\Large\bf
Wave function of the radion in the brane background
with a massless scalar field and a self-tuning problem
}
\author{$^{(a)}$Jihn E. Kim\footnote{jekim@phyp.snu.ac.kr}, $^{(b)}$Bumseok
Kyae\footnote{bkyae@bartol.udel.edu} and $^{(a)}$Hyun Min
Lee\footnote{minlee@phya.snu.ac.kr}}
\address{ $^{(a)}$Department of Physics and Center for Theoretical
Physics, Seoul National University, Seoul 151-747, Korea\\
$^{(b)}$Bartol Research Institute, University of Delaware, Newark,
DE 19716, USA}
\maketitle

\begin{abstract}
We consider flat solutions in the brane background with a massless
scalar field appearing in 5D $H^2_{MNPQ}$.  Since there exist bulk
singularities or arises the divergent 4D Planck mass, we should
introduce a compact extra dimension, the size of which is then
fixed by brane tension(s) and a bulk cosmological constant.
Inspecting scalar perturbations around the flat solutions, we find
that the flat solutions are stable vacua from the positive mass
spectrum of radion. We show that the massless radion mode is
projected out by the boundary condition arising in cutting off the
extra dimension. Thus, the fixed extra dimension is not alterable,
which is not useful toward a self-tuning of the cosmological
constant.
\end{abstract}


\newpage
\section{Introduction}
The Randall-Sundrum(RS) models\cite{rs1,rs2} have been initially motivated
for solving the
hierarchy problem in a geometric way\cite{rs1} and giving an alternative
to the conventional Kaluza-Klein compactification of the extra dimension
through the normalizable zero mode of graviton\cite{rs2}. On the other hand,
the fact that there exists the 4D flat solution in the RS models even
with nonzero brane tension(s) and a nonzero negative bulk cosmological
constant reminds us of the cosmological constant problem in a new
perspective and
thus the RS models deserve further investigations for searching a higher
dimensional solution to the cosmological constant
problem\cite{shapo,kachru,nilles,tye,tamvakis,holdom,kklcc,kklcc1,haya,flan}.
The RS solutions requires one (or two) fine-tuning conditions between brane
tension(s) and a bulk cosmological constant by consistency. Therefore,
if one loosen the fine-tuning condition(s), our universe should appear
as being curved.
Therefore, we should seek for models of evading fine-tuning parameters
with the requirements for the consistent solution to the cosmological
constant problem such that \\
\indent $\bullet$ there should be no naked singularities and \\
\indent $\bullet$ the 4D Planck mass should be finite for gravity
to be relevant.

Early attempts with a massless bulk scalar field coupled to the brane
in the flat bulk ended with
the naked singularity and reproduced a fine-tuning in regularizing the naked
singularity with another brane\cite{kachru,nilles}.
The main motivation for introducing a massless scalar field is to use the shift
symmetry of the scalar field to self-tune the cosmological constant
with its special coupling to the brane.

On the other hand, there have appeared suggestions of introducing
a massless bulk scalar field not coupled to the
brane\cite{holdom,kklcc1,haya}. In those attempts, the scalar field
just plays a role of bulk source only and thus it gives rise to
only a non-multiplicative integration constant of the warp factor
by making bulk solutions non-trivial. [The integration constant
from scalar field itself is not important since it is just
contained in the warp factor as multiplicative.] With $1/H^2$
term, where $H^2=H_{MNPQ}H^{MNPQ}$, it has been shown that there
exists a self-tuning solution\cite{kklcc}.

However, attempts with a conventional kinetic term of the scalar
field did not give a self-tuning solution, i.e. there appear naked
singularities or the warp factor becomes divergent away from the
brane. Therefore, it is necessary to introduce another brane for
cutting off the extra dimension, from which we concluded that
there does not exists a self-tuning solution when the scalar couples
to the brane~\cite{nilles}.
However, when the scalar does not couple to the brane, 
another brane introduces a parameter the
distance between two branes, as a result of which it seems that
there does not arise any direct fine-tunings but the extra
dimensional distance is determined by brane tensions and bulk
cosmological constant. In this case, there arises a question
whether the size of the extra dimension can be regarded as a kind
of integration constant to be determined by the boundary
condition.

In this paper, we study this question whether the extra
dimensional distance can be regarded as an integration constant or
not by investigating scalar perturbations around the flat
solutions obtained for arbitrary bulk cosmological constant. We
find that the mass spectrum of the radion is given as positive
definite and a massless mode is decoupled by the boundary
condition appearing in cutting off the extra dimension. Thus the
flat solutions are shown to be true minima of the action in taking
into account the boundary condition.

The absence of massless radion means that the size of extra
dimension is fixed but not alterable. Therefore, since the size of
extra dimension does not correspond to a kind of integration
constant in the self-tuning models, one cannot evade one
fine-tuning condition between brane tension(s) and bulk
cosmological constant for obtaining a flat solution with a
conventional kinetic energy term.

Using the remaining gauge transformations preserving
the scalar perturbations, we also find that there exist a massless graviscalar
and a massless spin-2 graviton while there is no massless vector mode.
The massive excitations of spin-2 graviton are shown to be positive,
which guarantees the stability of the flat solutions again.

This paper is organized as follows. In the next section, we
provide the model setup for consideration and present flat
solutions. In Sec. III, we perturb the scalar field around the
flat solutions and identify the radion spectrum. Then, in Sec. IV,
we present the graviton perturbations for completeness. Sec. V is
a conclusion.

\section{Model Setup}
On top of the RS model\cite{rs1,rs2},
we introduce a three form field in the bulk
without coupling to the brane. For future convenience of cutting off
the extra dimension, we include the sum of brane actions.
Then, the 5D action of our model setup is
\begin{equation} \label{action1}
S=\int d^4x\int dy \sqrt{-g}\left(\frac{M^3}{2}R-\Lambda_b
-\frac{1}{2\cdot 4!}H_{MNPQ}H^{MNPQ}+\sum_{i} {\cal L}_m^{(i)}\delta(y-y_i)
\right).
\end{equation}
In order to get 4D flat solutions, let us take the ansatz for the metric as
\begin{equation}
ds^2=\beta^2(y)\eta_{\mu\nu}dx^\mu dx^\nu+dy^2 \label{metric}
\end{equation}
where $(\eta_{\mu\nu})={\rm diag.}(-1,+1,+1,+1)$.
Then Einstein tensors are,
\begin{eqnarray}
G_{\mu\nu}&=&g_{\mu\nu}\left[3\left(\frac{\beta^\prime}{\beta}\right)^2
+3\left(\frac{\beta^{\prime\prime}}{\beta}\right)\right],\nonumber\\
G_{55}&=&6\left(\frac{\beta^\prime}{\beta}\right)^2.
\end{eqnarray}
where prime denotes differentiation with respect to $y$.
With the brane tension $\Lambda_1$ and $\Lambda_2$ at the $y=0$ and $y=y_c$
branes, respectively, and the bulk
cosmological constant $\Lambda_b$, the energy momentum tensors are
\begin{eqnarray}
T_{MN}&=& -g_{MN}\Lambda_b-\frac{\sqrt{-g^{(4)}}}{\sqrt{-g}}g_{\mu\nu}
\delta_M^\mu\delta_N^\nu
\sum_i\Lambda_i \delta(y-y_i)+\tilde T_{MN},\label{emtensor}\\
\tilde T_{MN}&=&\frac{1}{4!}\left(4H_{MPQR}H_N\,^{PQR}
-\frac{1}{2}H^2 g_{MN}\right) \label{4form}\\
&=&\nabla_M\phi\nabla_N\phi-\frac{1}{2}g_{MN}(\nabla\phi)^2,\label{scalar}
\end{eqnarray}
where we used the fact that a three form field in 5D spacetime is dual
to a scalar field as $H_{MNPQ}=\sqrt{-g}\,\epsilon_{MNPQ}\,^R\, \nabla_R\phi$.
Here when the dual relation is inserted in the kinetic term of the action,
its overall sign looks opposite compared
with the case of a scalar field but there does not arise
such inconsistency between the action and the Einstein's equation
when the surface term for the three form field is included.

We also take the ansatz for the nonvanishing components of four form field
$H_{\mu\nu\rho\sigma}$ as
\begin{equation}
H_{\mu\nu\rho\sigma}=\sqrt{-g}\,\epsilon_{\mu\nu\rho\sigma}f(y)
\label{g4}
\end{equation}
where $\mu,\cdots$ run over the Minkowski indices 0, 1, 2, and 3. With
the above ansatz, the field equation for the four form field is
satisfied,
\begin{equation}
\partial_{M}\bigg[\sqrt{-g}H^{MNPQ}\bigg]=0.
\end{equation}
The two relevant Einstein equations are the (55) and ($\mu\mu$) components,
\begin{eqnarray}
6\left(\frac{\beta^\prime}{\beta}\right)^2&=&-\Lambda_b
+\frac{A}{\beta^8}\label{eqnb}\\
3\left(\frac{\beta^\prime}{\beta}\right)^2+3\left(
\frac{\beta^{\prime\prime}}{\beta}\right)&=&
-\Lambda_b-\Lambda_1\delta(y)-\Lambda_2\delta(y-y_c)
-\frac{A}{\beta^8}\label{eqnb1}
\end{eqnarray}
where $A/\beta^8\equiv f^2/2$ expressed in terms of a `positive'
constant $A$. The solutions of Eq.~(\ref{eqnb}) and (\ref{eqnb1})
with $Z_2$ symmetry are\cite{kklcc1}
\begin{eqnarray}
{\rm for}~~~\Lambda_b<0~~:~~
(1)\, \beta(|y|)&=&\left(\frac{a}{k}\right)^{1/4} {[\sinh(-4k|y|+c)]^{1/4}}
\label{nega1}\\
(2)\, \beta(|y|)&=&\left(\frac{a}{k}\right)^{1/4} {[\sinh(4k|y|+c)]^{1/4}}
\label{nega2}\\
{\rm for}~~~\Lambda_b>0~~:~~
(1)\, \beta(|y|)&=&\left(\frac{a}{k}\right)^{1/4} {[\sin(-4k|y|+c)]^{1/4}}
\label{posi1}\\
(2)\, \beta(|y|)&=&\left(\frac{a}{k}\right)^{1/4} {[\cos(-4k|y|+c)]^{1/4}}
\label{posi2}\\
{\rm for}~~~\Lambda_b=0~~:~~
(1)\, \beta(|y|)&=&\left(-4a|y|+c\right)^{1/4}, \label{zero1}\\
(2)\, \beta(|y|)&=&\left(4a|y|+c\right)^{1/4}, \label{zero2}
\end{eqnarray}
where $k\equiv\sqrt{|\Lambda_b|/6}$ and the $a$ is defined in terms of $A$,
\begin{equation}
a\equiv \sqrt{\frac{A}{6}}.\label{a}
\end{equation}
We note that $\beta$'s of Eqs.~(\ref{nega2}) and (\ref{zero2}) do not give
localized gravity on the $y=0$ brane while $\beta$'s of Eqs.~(\ref{nega1}),
(\ref{posi1}), (\ref{zero1}) have naked singularities
at $|y|=c/(4k)$ or $|y|=c/(4a)$ and $\beta$ of Eq.~(\ref{posi2}) does at
$|y|=(c+\pi/2)/(4k)$.
Therefore, to get the effective four dimensional gravity or to avoid the
sigularities in the bulk, it is indispensable to cut the extra dimension
such that it has a finite length size by introducing another brane.

Then, since the $Z_2$ symmetry and the periodicity for the compact dimension
gives rise to the boundary conditions at the branes,
\begin{equation}
\frac{\beta^\prime}{\beta}\Big|_{y=y_i^+}\equiv -\frac{\Lambda_i}{6},\label{bc1}
\end{equation}
the consistency requires the following relations for the above three cases
\begin{eqnarray}
{\rm for}~~~\Lambda_b<0~~:~~
\pm c&=&{\rm coth}^{-1}\left(\frac{k_1}{k}\right)
=4ky_c-{\rm coth}^{-1}\left(\frac{k_2}{k}\right) \label{tune1}\\
{\rm for}~~~\Lambda_b>0~~:~~
(1)\, c&=&{\rm cot}^{-1}\left(\frac{k_1}{k}\right)
=4ky_c-{\rm cot}^{-1}\left(\frac{k_2}{k}\right) \label{tune2}\\
(2)\, c&=&-{\rm tan}^{-1}\left(\frac{k_1}{k}\right)
=4ky_c+{\rm tan}^{-1}\left(\frac{k_2}{k}\right) \label{tune3}\\
{\rm for}~~~\Lambda_b=0~~:~~
\pm c&=&\frac{a}{k_1}=a\left(4y_c-\frac{1}{k_2}\right),\label{tune4}
\end{eqnarray}
where $c>0$ for $\Lambda_b\leq 0$ and $0<c<\pi/2$ for $\Lambda_b>0$ and
\begin{equation}
k_1\equiv \frac{\Lambda_1}{6}, \, \,\, k_2\equiv \frac{\Lambda_2}{6}.
\end{equation}
For all the above cases, there do not appear any direct
fine-tuning conditions between brane tensions unlike the RS case
but the size of the extra dimension is determined by the brane
tensions and the bulk cosmological constant. In particular, for
the $\Lambda_b>0$ case (Eq.~(\ref{posi2})), as recently argued in
Ref.\cite{haya}, it may be possible to compactify the extra
dimension without the need of introducing another brane by
identifying the two extrema of the warp factor symmetric around
the $y=0$ brane. In Fig. 1 we show a schematic behavior of the
warp factor. If Point Q corresponds to the location of the another
brane, then one has to introduce a brane tension there. However, if
Point P is chosen, it is not necessary to introduce another brane.
That means the extra dimensional size is up to Point P which is
determined by the other parameters in the theory. For this to be
the case, we must have a shift symmetry in the radion direction
whose distance is determined by the parameters in the theory.
Namely, we only have to take the boundary condition as the first
one in Eq.~(\ref{tune3}) and the length size of the extra
dimension can be regarded as being determined by the brane tension
as $y_c=c/(4k)=-(4k)^{-1}{\rm tan}^{-1}(k_1/k)$ for $k_1<0$.

\vskip 0.3cm
\begin{figure}[b]
\centering \centerline{\epsfig{file=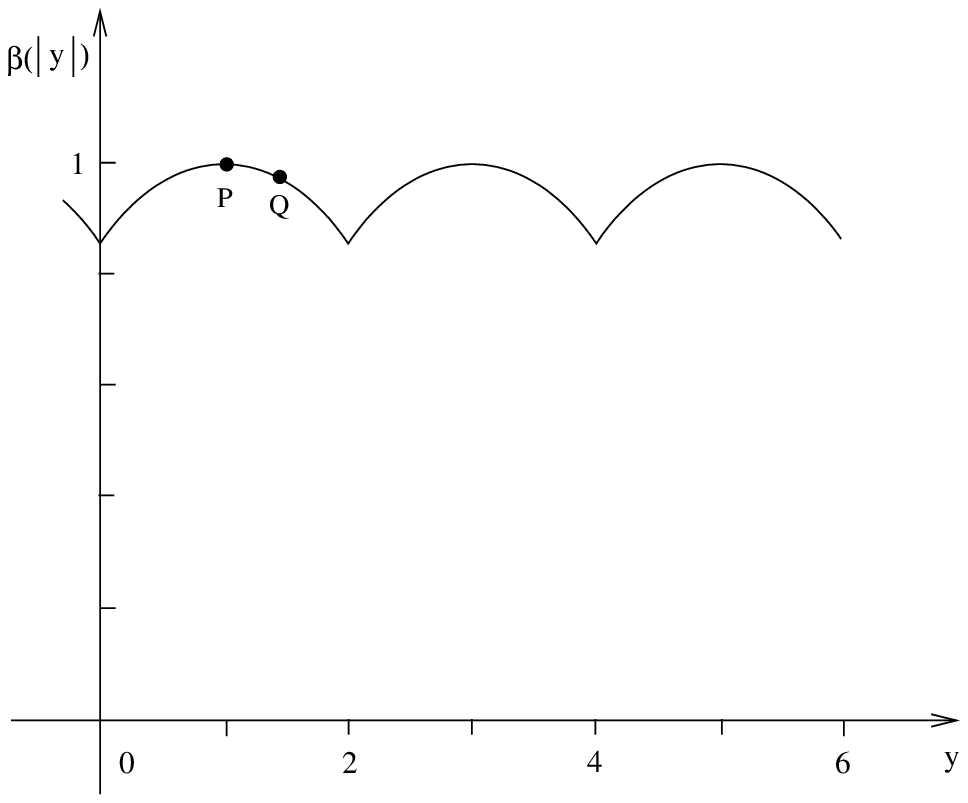,width=90mm}}
\end{figure}
\centerline{ Fig.~1.\ \it  The schematic behavior of the warp
factor for $\Lambda_b>0$.}
\vskip -0.2cm
\centerline{\it Point P
is the case of our main concern.} \vskip 0.3cm

\section{Scalar perturbations and radion}
We expect that there exists the radion mode as a perturbation around the 4D
flat solutions obtained, which corresponds to altering
the positions of the branes or the size of the extra dimension and appears a
scalar field in the effective 4D theory. Thus let us make scalar perturbations
around the flat solutions for investigating the radion spectrum in our model.

In the RS case without radius stabilization, there exists a massless radion
for the flat solution since the size of extra dimension is not
determined\cite{rs1,terning,ruba}.
And it is also shown that
the radion has a positive mass for the 4D AdS solution
while it has a negative mass for the 4D dS solution\cite{chacko,binetruy}.
For the stabilized RS model, the radion mass is also investigated in detail
in Ref.\cite{csaki}. In models with metastable graviton and multigravity
scenarios, it is shown that the radion dynamics is crucial to test
their stability\cite{pilo,kogan}.

As we mentioned in the previous section,
we regard the three form field as a scalar field by duality
on investigating perturbations for convenience. In that case, we only have
to deal with the energy momentum tensor coming from the scalar field
as in Eq.~(\ref{scalar}) and the scalar field equation such as
\begin{equation}
\partial_M(\sqrt{-g}\partial^M \phi)=0.
\end{equation}
We thus take a general ansatz for the metric perturbations as
\begin{eqnarray}
ds^2&=&K^{-2}(z)(\eta_{MN}+h_{MN}(x,z))dx^M dx^N
\label{gmpert}\\
\phi(x,z)&=&\phi_0(z)+\varphi(x,z),\label{scalarpert}
\end{eqnarray}
where $z=\int^z dy/\beta(y)$, $K(z)=1/\beta(y(z))$ and $(\phi'_0)^2=AK^6$.

Then, the linearized Einstein's tensor, the linearized energy momentum
tensor and the linearized scalar equation read
\begin{eqnarray}
\delta G_{MN}&=&-\frac{\Box}{2} \bar{h}_{MN}
+\partial_{(M}\partial^P\bar{h}_{N)P}
-\frac{1}{2}\eta_{MN}\partial^P\partial^Q\bar{h}_{PQ}
-\frac{3K'}{2K}(\partial_Mh_{N5}+\partial_Nh_{M5}
-\partial_5h_{MN}) \nonumber\\
&&-3\eta_{MN}\bigg[\left(-\frac{K''}{K}
+2\frac{K'^2}{K^2}\right)h_{55}
-\frac{K'}{K}\partial^P\bar{h}_{P5}\bigg]
-3\bigg[\frac{K''}{K}-2\frac{K'^2}{K^2}\bigg]h_{MN},\label{lineq}\\
\delta T_{\mu \nu}&=&\left(-3\frac{K''}{K}+
6\frac{K'^2}{K^2}\right)h_{\mu \nu}
+\sum_i\delta(z-z_i)\frac{\Lambda_i}{2K}h_{55}
\eta_{\mu \nu}
+\frac{1}{2}(\phi'_0)^2 h_{55}\eta_{\mu \nu}-\phi'_0 \varphi'\eta_{\mu\nu}, \\
\delta T_{\mu 5}&=&6\frac{K'^2}{K^2}h_{\mu 5}+\phi'_0\partial_\mu\varphi,\\
\delta T_{5 5}&=&6\frac{K'^2}{K^2}h_{55}-\frac{1}{2}(\phi'_0)^2 h_{55}
+\phi'_0\varphi',\label{lineqend}
\end{eqnarray}
\begin{eqnarray}
\Box\varphi
-3\frac{K'}{K}\varphi'+\frac{1}{2}\phi'_0 (h^{\mu\prime}_\mu-h'_{55})=0,
\label{linsc}
\end{eqnarray}
where $(M,N)$ is a half of the symmetric combination, $'$ denotes
the derivative with respect to $z$ and $\partial^M\equiv
\eta^{MN}\partial_N$, $\Box\equiv \eta^{\mu
\nu}\partial_{\mu}\partial_{\nu} +\partial_z^2$. $\bar{h}_{MN}$ is
defined as $\bar{h}_{MN}\equiv h_{MN}-\eta_{MN}h/2$ and $h$ is the
trace for $h_{MN}$.

Since we are interested in the scalar perturbations for investigating
the radion, let us take a gauge choice in the metric ansatz (\ref{gmpert}) as
\begin{equation}
h_{\mu\nu}(x,z)=F(x,z)\eta_{\mu\nu},~h_{\mu 5}=0,~h_{55}(x,z)=G(x,z).
\label{mpert}
\end{equation}
Thus, the ($\mu\nu$) component of the linearized Einstein tensor is
written as the form
\begin{equation}
\delta G_{\mu\nu}=\partial_\mu\partial_\nu \bigg(-F-\frac{1}{2}G
\bigg) +\cdot\cdot\cdot
\end{equation}
where the ellipses all contain terms proportional to $\eta_{\mu\nu}$. And the
linear perturbations from the energy momentum tensor are also
$\sim \eta_{\mu\nu}$. Therefore, we obtain an immediate relation
such as
\begin{equation}
G=-2F.\label{fix1}
\end{equation}
Then the linearized ($\mu\mu$), ($\mu 5$) and (55) Einstein's equations
and the linearized scalar field equation are
\begin{equation}
\frac{3}{2}F^{\prime\prime}-\frac{15}{2}\frac{K'}{K}F'
+6\bigg[2\bigg(\frac{K'}{K}\bigg)^2-\frac{K^{\prime\prime}}{K}\bigg]F
+\bigg[\sum_i\frac{\Lambda_i}{K}\delta(z-z_i)+(\phi'_0)^2\bigg]F
+\phi'_0\varphi'=0,
\label{pert1}
\end{equation}
\begin{equation}
-\frac{3}{2}\partial_\mu F'-3\frac{K'}{K}\partial_\mu F
=\phi'_0\partial_\mu\varphi \label{pert2}
\end{equation}
\begin{equation}
\frac{3}{2}\partial^\mu\partial_\mu F-6\frac{K'}{K}F'
+\bigg[12\bigg(\frac{K'}{K}\bigg)^2-(\phi'_0)^2\bigg]F
-\phi'_0\varphi'=0 \label{pert3}
\end{equation}
\begin{equation}
\partial^\mu\partial_\mu\varphi+\varphi^{\prime\prime}
-3\frac{K'}{K}\varphi'+3\phi'_0 F'=0.\label{scpert}
\end{equation}
The ($\mu 5$) component equation (\ref{pert2}) can be integrated to give
\begin{equation}
\phi'_0\varphi=-\frac{3}{2}\bigg(F'-2\frac{K'}{K}F\bigg)
+f(z).\label{fix2}
\end{equation}
The metric ansatz (\ref{mpert}), Eqs.~(\ref{fix1}) and
(\ref{fix2}) with $f(z)=0$ fix our gauge choice. Eliminating
$\phi'_0\varphi'$ from Eqs.~(\ref{pert1}) and (\ref{pert3}) gives
rise to the equation for $F$ only as
\begin{equation}
\partial^\mu\partial_\mu F+F^{\prime\prime}-9\frac{K'}{K}F'
+\bigg[16\bigg(\frac{K'}{K}\bigg)^2 -4\frac{K^{\prime\prime}}{K}
+\sum_i\frac{2\Lambda_i}{3K}\delta(z-z_i)\bigg]F=0
\end{equation}
from which we can easily obtain the boundary conditions at the branes
\begin{equation}
\bigg[F'-2\frac{K'}{K}F\bigg]_{z=z_i^+}=0,\label{radbc}
\end{equation}
where we used
$K^{\prime\prime}/K=2AK^6/3+(\Lambda_i/3K)\delta(z-z_i)$ and
$\Lambda_i=6K'|_{z=z_i}$. It is staightforward to check the scalar
field equation (\ref{scpert}) by multiplying Eq.~(\ref{scpert})
with $\phi'_0$ and using Eqs.~(\ref{fix2}), (\ref{pert3}) and the
background equations (\ref{eqnb}), (\ref{eqnb1}) in the $z$
coordinate. Thus, $F$ must satisfy Eq.~(\ref{radbc}) and the bulk
equation
\begin{equation}
\partial^\mu\partial_\mu F+F^{\prime\prime}-9\frac{K'}{K}F'
+\bigg[16\bigg(\frac{K'}{K}\bigg)^2
-\frac{4K^{\prime\prime}}{K}\bigg]F=0.
\end{equation}
Note that studying $F$ is equivalent to studying the radion, in
view of Eqs. (\ref{mpert}) and (\ref{fix1}).

To make a Kaluza-Klein reduction of the radion field to 4D, we choose
a separation of variables as $F(x,z)=\psi(z)\rho(x)$.
Taking a rescaling such as $\tilde\psi=\psi/(\phi'_0 K^{3/2})$\cite{gio},
we obtain the 4D equation of the radion field and the bulk equation
for the wave function of the radion
\begin{eqnarray}
(\partial^\mu \partial_\mu-m^2)\rho(x)&=&0,\label{4deq}\\
(-\partial^2_z+V(z))\tilde\psi(z)&=&m^2\tilde\psi(z)\label{bulkeq}
\end{eqnarray}
where
\begin{eqnarray}
V(z)&=&\xi\bigg(\frac{1}{\xi}\bigg)^{\prime\prime}, \\
\xi&=&\frac{\phi'_0}{K^{1/2}K'}.
\end{eqnarray}
The above equation corresponds to nothing but a supersymmetric quantum
mechanics
\begin{eqnarray}
Q^\dagger Q\tilde\psi(z)\equiv\bigg[\partial_z
+\xi\bigg(\frac{1}{\xi}\bigg)'\bigg]
\bigg[-\partial_z+\xi\bigg(\frac{1}{\xi}\bigg)'\bigg]
\tilde\psi(z)=m^2\tilde\psi(z).
\end{eqnarray}
The hermitianity of the above differential operator guarantees the positivity
of the mass spectrum with $m^2\geq 0$ : there is no tachyonic mode of radion.
The bulk solution for the massless radion field with $m^2=0$ is given
by a linear combination such as
\begin{equation}
\tilde\psi_0(z)=\frac{1}{\xi}(c_0+d_0\int^z_0 dz\ \xi^2)\label{massless0}
\end{equation}
or
\begin{equation}
\psi_0(z)=K^2 K'\bigg[c_0+d_0\int^z_0\,dz
\bigg(\frac{\phi'_0}{K^{1/2}K'}\bigg)^2\bigg]\label{massless}
\end{equation}
where $c_0$ and $d_0$ are integration constants.
Then, we can take the wave function of the massless radion to be
consistent with the $Z_2$ symmetry
\begin{eqnarray}
\psi_0(z)=\left\{ \begin{array}{l} K^2 K'\bigg[c_0+d_0\int^z_0\,dz
\bigg(\frac{\phi'_0}{K^{1/2}K'}\bigg)^2\bigg] ~~{\rm for}~~ z>0,\\
\\
-K^2 K'\bigg[c_0-d_0\int^z_0\,dz
\bigg(\frac{\phi'_0}{K^{1/2}K'}\bigg)^2\bigg] ~~{\rm for}~~ z<0.
\end{array}\right.
\end{eqnarray}
Then, using the boundary condition at the $z=0$ (or $y=0$) brane
from Eq.~(\ref{radbc}),
\begin{equation}
\bigg[\psi'-2\frac{K'}{K}\psi\bigg]_{z=0^+}=0,
\end{equation}
we obtain the ratio between two integration constants as
\begin{equation}
{\rm (A)}~~:~~\frac{c_0}{d_0}
=-\bigg[\frac{(\phi'_0)^2}{K^{\prime\prime}K' K}\bigg]_{z=0^+}.
\label{ratio1}
\end{equation}
On the other hand, since
the extra dimension should be cut off with or without another brane
to escape a bulk singularity or a divergent 4D Planck mass,
an additional boundary condition appears.
For the case without another brane, the derivative of the metric perturbation
should be zero at the end of the extra dimension as for the background metric.
Therefore, in view of Eq.~(\ref{radbc}), another boundary condition should be
the same irrespective of the existence of an additional brane as
\begin{equation}
(K^{-2}\psi)'\Big|_{z=z_c^-}=0.
\end{equation}
Thus we also find another necessary condition for integration constants
\begin{equation}
{\rm (B)}~~:~~
\frac{c_0}{d_0}=-\bigg[\frac{(\phi'_0)^2}{K^{\prime\prime}K'
K}\bigg]_{z=z^-_c}-\int^{z_c}_0 dz
\bigg(\frac{\phi'_0}{K^{1/2}K'}\bigg)^2.\label{ratio2}
\end{equation}
However, the two equations (\ref{ratio1}) and (\ref{ratio2})
cannot be satisfied simultaneously because the difference
$(A)-(B)$ between the two turns out to be nonzero
\begin{equation}
(A)-(B)=-3\int^{z_c}_0 dz K^{-3}\ne 0.\label{differ}
\end{equation}
As a
result, {\it the massless radion mode should be regarded as being
projected out} by another boundary condition appearing in cutting
off the extra dimension, irrespective of whether the extra
dimension is compactified with one or two branes.

{\it The absence of massless radion implies that it is impossible
to change the brane spacing} or the size of extra dimension for
the flat solution to exist for different sets of brane tension(s)
and bulk cosmological constant. Thus the size of extra dimension
does not correspond to an integration constant in the self-tuning
models and one cannot escape the one fine-tuning
condition\cite{kklcc1} between input parameters for obtaining a
flat solution for initially fixed extra dimension as seen in
Eqs.~(\ref{tune1})-(\ref{tune4}).

On the other hand, massive radion modes are alive with discrete
positive masses since there appear generically two integration
constants and mass to be determined : one integration constant and
mass are determined by two boundary conditions at the ends of the
extra dimension and the other integration constant is fixed by
normalization of the wave function of the radion. Therefore, the
size of the extra dimension fixed by
Eqs.~(\ref{tune1})-(\ref{tune4}) is stable under perturbations,
which guarantees that the flat solutions of the Einstein's
equations Eqs.~(\ref{nega1})-(\ref{zero2}) are true minima of the
action.

\section{graviton perturbations}
Even though we made a gauge choice such that only scalar perturbations are
taken into account in the previous section, there should be the remaining gauge
transformations preserving the gauge conditions (\ref{fix1})
and (\ref{fix2}) with the metric ansatz (\ref{mpert}),
which enables us to impose a convenient 4D gauge for the
graviton.
Thus, for completeness, on top of the scalar perturbations corresponding
to the radion in the previous section, let us consider the graviton
perturbations with the general metric ansatz
\begin{equation}
h_{\mu\nu}(x,z)=F(x,z)\eta_{\mu\nu}+\tilde{h}_{\mu\nu}(x,z),~
h_{\mu 5}\neq 0,~ h_{55}(x,z)=G(x,z).\label{geng}
\end{equation}
Then, with the gauge conditions (\ref{fix1}) and (\ref{fix2})
for the scalar perturbations, we use a 4D gauge transformations to impose
the de Donder gauge condition for the graviton
\begin{equation}
0=\partial^\mu \bar{\tilde{h}}_{\mu\nu}=\partial^\mu \bar{h}_{\mu\nu}
\label{deD}
\end{equation}
where $\bar{h}_{\mu\nu}\equiv h_{\mu\nu}-\frac{1}{2}h^\lambda_\lambda
\eta_{\mu\nu}$
and the second equality comes from Eq.~(\ref{fix1}). Now we have fixed five
of the fifteen degrees of freedom of the 5D metric perturbations.

Assuming that the equations for scalar perturbations
(\ref{pert1})-(\ref{scpert}) are reproduced for the general gauge choice
with Eqs.~(\ref{geng}) and (\ref{deD}), we obtain
the equations for the graviton from (\ref{lineq})-(\ref{lineqend}) as
$$
-\frac{1}{2}\bigg(\Box-3\frac{K'}{K'}\partial_5\bigg)\tilde{h}_{\mu\nu}
+\frac{1}{4}\eta_{\mu\nu} \bigg(\Box+\partial^2_5
-6\frac{K'}{K}\partial_5\bigg)\tilde{h}^\mu_\mu
$$
\begin{equation}
+\bigg(\partial_5-3\frac{K'}{K}\bigg)(\partial_{(\mu} h_{\nu)5}
-\eta_{\mu\nu}\partial^\lambda h_{\lambda 5})=0,\label{res1}
\end{equation}
\begin{equation}
-\frac{1}{2}\bigg(\partial^\lambda\partial_\lambda
+6\frac{K^{\prime\prime}}{K}\bigg)h_{\mu 5}
+\frac{1}{2}\partial_\mu \bigg(\partial^\lambda h_{5\lambda}
-\frac{1}{2}\partial_5 \tilde{h}^\lambda_\lambda\bigg)=0,\label{res2}
\end{equation}
\begin{equation}
\frac{1}{4}\partial^\lambda\partial_\lambda\tilde{h}^\mu_\mu
+\frac{3K'}{K}\bigg(\partial^\lambda h_{\lambda 5}-\frac{1}{2}\partial_5
\tilde{h}^\mu_\mu\bigg)=0.\label{res3}
\end{equation}
Then we find that both the trace of Eq.~(\ref{res1}) and Eq.~(\ref{res3})
can be satisfied simultaneously with
\begin{equation}
\partial^\lambda h_{5\lambda}-\frac{1}{2}\partial_5 \tilde{h}^\mu_\mu=0,
\label{decouple}
\end{equation}
under which Eqs.~(\ref{res3}) and (\ref{res2}) become, respectively,
\begin{equation}
\partial^\lambda\partial_\lambda \tilde{h}^\mu_\mu=0,\label{trace}
\end{equation}
\begin{equation}
\bigg(\partial^\lambda\partial_\lambda
+6\frac{K^{\prime\prime}}{K}\bigg)h_{\mu 5}=0.\label{vector}
\end{equation}
Therefore, there appears a massless gravi-scalar field
$\tilde{h}^\mu_\mu$, which couples to the trace of the 4D energy
momentum tensor as for the radion. On the other hand, since
Eq.~(\ref{vector}) cannot be satisfied either on the brane or in
the bulk, $h_{\mu 5}=0$, i.e., there is no massless vector mode,
which is consistent with Eq.~(\ref{decouple}) since
$(\tilde{h}^\mu_\mu)'=0$ from comparing the linearized scalar
equations (\ref{linsc}) and (\ref{scpert}).

>From Eq.~(\ref{res1}), we also observe that the transverse traceless
spin-2 graviton ($\tilde{h}^{TT}_{\mu\nu}$) is automatically
decoupled due to $\tilde{h}^{\mu \prime}_\mu=h_{\mu 5}=0$ and
the de Donder gauge (\ref{deD}):
\begin{equation}
\bigg(\Box-3\frac{K'}{K}\partial_5\bigg)\tilde{h}^{TT}_{\mu\nu}=0.
\end{equation}
Consequently, it turns out that there exists a massless spin-2 graviton
since the above equation is satisfied by
$\tilde{h}^{TT}_{\mu\nu}(x,z)=c_0 e^{ipx}\epsilon_{\mu\nu}$
with $p^2=0$, where $c_0$ is a constant
and $\epsilon_{\mu\nu}$ is a polarization tensor.

With a separation of variables as
$\tilde{h}^{TT}_{\mu \nu}=K^{3/2}(z)\tilde{\psi}(z)e^{ipx}\epsilon_{\mu \nu}$
$(p^2=-m^2)$, we obtain a Schr\"{o}dinger-like equation again
\begin{equation}
\bigg(-\partial_z^2+V(z)\bigg)\tilde{\psi}(z)=m^2\tilde{\psi}(z)
\end{equation}
where
\begin{equation}
V(z)=\frac{15}{4}\bigg(\frac{K'}{K}\bigg)^2-\frac{3K^{\prime\prime}}{2K}.
\end{equation}
Therefore, we find that $m^2\geq 0$, i.e., there is no tachyonic state of the
graviton since the above equation can be regarded
as a supersymmetric quantum mechanics
\begin{eqnarray}
Q^\dagger Q\tilde{\psi}(z)\equiv
\bigg(\partial_z-\frac{3}{2}\frac{K'}{K}\bigg)
\bigg(-\partial_z-\frac{3}{2}\frac{K'}{K}\bigg)\tilde{\psi}(z)
=m^2\tilde{\psi}(z).
\end{eqnarray}

\section{Conclusion}
In this paper, by introducing a 5D massless scalar field not coupled to the
brane, we found that there exist flat solutions without direct fine-tunings
between brane tension(s) and a bulk cosmological constant.
However, since there arises a naked singularity or the 4D Planck
mass becomes divergent for non-compact extra dimension,
the extra dimension should be compactified with or without another brane
and then the size of extra dimension is fixed by brane and bulk
cosmological constants.

At first sight, the size of extra dimension looks a kind of
integration constant for the existence of flat solutions without
fine-tunings. However, on inspecting scalar perturbations around
the flat solutions, the absence of massless radion mode implies
that it is impossible to use the brane spacing as an integration
constant and the self-tuning of the cosmological constant is not
realized with the radion.

For completeness, using the remaining gauge transformations
preserving the scalar perturbations, we find that there exist a
massless gravi-scalar and a massless spin-2 graviton while there
is no massless vector mode. The spectrum of massive spin-2
graviton is shown to be non-negative, which implies the stability
of the flat solutions with a massless bulk scalar field together
with the positivity of the radion spectrum.

\acknowledgments This work is supported in part by the BK21
program of Ministry of Education(JEK, HML), the KOSEF Sundo
Grant(JEK, HML), and by the Center for High Energy Physics(CHEP),
Kyungpook National University(JEK, HML), and by US Department of
Energy DE-FG02-91ER40626(BK).

\end{document}